\documentclass[12pt]{iopart}

\usepackage{graphicx}
\usepackage{multirow}
\usepackage{cite}
\include{setstack.sty}

\begin{document}
\title[Exchange interactions in transition metal oxides]{Exchange interactions in transition metal oxides: The role of oxygen spin polarization}

\author{R~Logemann$^1$, A~N~Rudenko$^1$, M~I~Katsnelson$^1$ and A~Kirilyuk$^1$}
\address{$^1$ Radboud University, Institute for Molecules and Materials, NL-6525 AJ Nijmegen, The Netherlands}

\ead{R.Logemann@science.ru.nl}
\begin{abstract}
Magnetism of transition metal (TM) oxides is usually described in terms of the Heisenberg model, with orientation-independent interactions between the spins. However, the applicability of such a model is not fully justified for TM oxides because spin polarization of oxygen is usually ignored. In the conventional model based on the Anderson principle, oxygen effects are considered as a property of the TM ion and only TM interactions are relevant. Here, we perform a systematic comparison between two approaches for spin polarization on oxygen in typical TM oxides. To this end, we calculate the exchange interactions in NiO, MnO, and hematite (Fe$_2$O$_3$) for different magnetic configurations using the magnetic force theorem. We consider the full spin Hamiltonian including oxygen sites, and also derive an effective model where the spin polarization on oxygen renormalizes the exchange interactions between TM sites. 
Surprisingly, the exchange interactions in NiO depend on the magnetic state if spin polarization on oxygen is neglected, resulting in non-Heisenberg behavior. In contrast, the inclusion of spin polarization in NiO makes the Heisenberg model more applicable. Just the opposite, MnO behaves as a Heisenberg magnet when oxygen spin polarization is neglected, but shows strong non-Heisenberg effects when spin polarization on oxygen is included. In hematite, both models result in non-Heisenberg behavior. General applicability of the magnetic force theorem as well as the Heisenberg model to TM oxides is discussed.
\end{abstract}
\pacs{71.70.Gm, 75.30.Et, 75.50.Ee}
\maketitle

\section{Introduction}
Transition metal (TM) oxides constitute an interesting class of materials with a wide variety of physical phenomena such as superconductivity, colossal magnetoresistance, ferroelectricity, metal-insulator transitions and molecular magnets.\cite{Goodenough1963, gubanov2012magnetism, Khomskii_2014, Mott1974, Imada1998}
Many of these phenomena are closely related to their magnetic properties. Furthermore, macroscopic magnetic properties such as the magnetic order, Curie temperature or the magnon dispersion in these materials, require understanding of the underlying microscopic interactions. 

One of the most common and successful microscopic models for magnetism is the Heisenberg model: $H = - \sum_{i,j} J_{ij} \bi{s}_i \cdot \bi{s}_j$.\cite{vonsovskii1974magnetism,Kei1996,White2007} Its basic assumption is the localized nature of the magnetic moments (spins). Determination of the model parameters, i.e., the interatomic exchange interactions $J_{ij}$ and the magnetic moments $\bi{s}_i$, was a very active topic in the last decades.  
Several methods exist for mapping the full single-particle Hamiltonian obtained from first-principles electronic structure calculations onto the Heisenberg model. One of the widely used approaches is to consider several magnetic configurations and use the calculated energies to approximate the interatomic exchange interactions. The disadvantage of this method is that the Heisenberg model is \emph{a priori} assumed, while the angular dependence of the exchange interactions and magnetic moments cannot be determined. This method is therefore insufficient to determine the applicability of the Heisenberg model in general. 
An alternative approach is the so-called magnetic force theorem (MFT), where the exchange interactions are considered in the limit of infinitesimal rotations of the spins and are then calculated via the second variation of the total energy using a single magnetic state. \cite{Liechtenstein1984,Lichtenstein1987,Katsnelson2000} The MFT proved to be a powerful method for studying magnetic interactions from first principles.\cite{Turek2001,Turek2006a,Katsnelson2008,Sato2010,Ebert2011} The necessity of only a single magnetic state allows to study the applicability of the Heisenberg model by determining the dependence of the exchange interactions on the magnetic states. 

The validity of the Heisenberg model has been tested before for a variety of systems. In bulk metals, the applicability of the Heisenberg model has been studied for bcc Fe and fcc Ni, and also for Mn impurities in both metals.\cite{Turzhevskii1990}  Whereas bcc Fe shows Heisenberg behavior at the ferromagnetic (FM) state, MnFe alloys and FM Ni exhibit strong non-Heisenberg exchange mechanisms. 
Besides the bulk systems, also finite systems such as the molecular magnets V$_{15}$\cite{Boukhvalov2004} and Mn$_{12}$\cite{Boukhvalov2002,Mazurenko2014} have been studied. For both V$_{15}$ and Mn$_{12}$, the exchange interactions between the antiferromagnetic (AFM)/ferrimagnetic and FM configuration differ by no more than 20-30$\%$, a typical accuracy of the calculated Heisenberg model parameters in general. 
Furthermore, higher order exchange interactions such as biquadratic and ring exchange have been shown to lead to non-Heisenberg behavior in perovskite manganites.\cite{Fedorova2015} In high $T_c$ superconductors the ring and biquadratic exchange contributions are also proposed to be crucial for the description of magnetism.\cite{Coldea2001,Wysocki2010}

Surprisingly, in AFM TM oxides the non-Heisenberg effects have never been studied systematically to our knowledge. Typically, only the equilibrium (AFM) ground state is considered in the calculations of exchange interactions within the MFT approach. The resulting exchange parameters are relevant, therefore, for truly Heisenberg magnets only or in the limit of small deviations of spins, which corresponds to the regime of low-lying excitations. At the same time, non-equilibrium magnetism, atomistic spin dynamics and excitations in finite systems such as surfaces or nanostructures, require one to go beyond those limitations.\cite{Boukhvalov2004,Skubic2008,Mazurenko2014,Mikhaylovskiy2015}

In order to extract the exchange interactions from first-principles calculations, one should first map the electronic Hamiltonian onto the Heisenberg model. This mapping is by itself not uniquely defined. In the conventional picture, the Heisenberg model for typical AFM TM oxides such as NiO, MnO and hematite (Fe$_{2}$O$_{3}$) involves TM sites only, whereas the oxygen atoms mediate the magnetic interaction via super- or/and double exchange mechanisms. However, from CrO$_{2}$ and pyroxenes it is known that for the FM state, spin polarization on oxygen occurs via $p$-$d$ hybridization. This spin polarization can drastically change the exchange interactions and stabilize FM exchange interactions.\cite{Streltsov2008,Solovyev2015} A similar mechanism is responsible for the higher-order non-Heisenberg effects observed in FeRh, being the result of the hybridization between Fe and Rh.\cite{Mryasov2005}

Summarizing, two basic options exist to construct the Heisenberg model for TM oxides. The first one is based on the Anderson principle and assumes that oxygen and ligand effects are a property of the TM, whereas induced spin polarization should be ignored.\cite{Anderson1959} In this approach, only exchange interactions between the TM sites are considered, while oxygen plays the role of a mediator of indirect (super- or double exchange) interactions. 
In the second option, oxygen is also considered as a magnetic center provided it is sufficiently polarized. As a result, additional exchange interactions between TM and oxygen sites may come into play. In turn, the extended model that includes oxygen sites can be mapped onto an effective model involving renormalized interactions between the TM sites.

In this work, we compare the performance of both approaches for three prototype TM oxides: NiO, MnO and hematite (Fe$_{2}$O$_{3}$). We calculate magnetic moments and exchange interactions starting from different magnetic configurations (FM and AFM) for the three materials. We derive an effective model to consider spin polarization on oxygen and show how the exchange interactions between TM sites are primarily affected. Finally, we perform a general comparison between the two approaches. We find that explicit treatment of oxygen spin polarization reduces the non-Heisenberg effects in NiO. That is, the exchange interactions depend on the magnetic state considered. In contrast, the inclusion of spin polarization in MnO makes the non-Heisenberg effects more pronounced. For hematite, both approaches result in non-Heisenberg behavior. 

The paper is organized as follows. In section~\ref{sec:method} the theoretical methods for the calculation of the electronic structure and exchange interactions are presented. Section~\ref{sec:resultsdiscussion} contains the results for NiO, MnO, and Fe$_{2}$O$_{3}$. In section~\ref{sec:disc}, we discuss the results and give a general comparison of the materials studied. Finally, the paper is concluded in section~\ref{sec:conclusion}. 

\section{\label{sec:method}Methods}
In this Section, we first provide details on the electronic structure calculations using density functional theory (DFT), and describe the mapping procedure (section~\ref{sec:method}~A). Next, we overview the MFT approach for the calculation of exchange interactions (section~\ref{sec:method}~B). We then present an alternative scheme based on the total energy calculations (section~\ref{sec:method}~C), and finally derive an effective Heisenberg model that captures the effects of induced spin polarization (section~\ref{sec:method}~D). 

\subsection{Electronic structure}
To calculate the electronic structure using DFT, we use the Vienna \emph{ab initio} simulation package (\textsc{vasp}),\cite{Kresse1996} which implements the projector augmented wave (PAW) method.\cite{Blochl1994,Kresse1999}
For NiO and MnO we used an undistorted fcc unit cell with the lattice constants $4.17$~\AA{} and $4.44$~\AA{}, respectively. 
A $\Gamma$-centered $12 \times 12 \times 12$ {\bf k}-point grid and an energy cutoff of 500~eV for the plane waves were used for both materials. To describe the exchange-correlation effects, we use the rotationally invariant PBE$+U$ functional as proposed by Dudarev\cite{Dudarev1998} for NiO and MnO. For NiO, we used the Hubbard parameters $U = 6.3$~eV and $J = 1$~eV, consistent with previous works.\cite{Dudarev1998,Zhang2006} For MnO, the calculations are performed using $U = 6.9$~eV and $J = 0.86$~eV.\cite{Franchini2005}
For hematite, we employ the rotationally invariant PBE$+U$ method as proposed by Liechtenstein\cite{Lichtenstein1995} with $U = 5$~eV and $J = 1$~eV, and use the crystallographic structure reported in \cite{Blake1966}, which we geometrically optimized for the equilibrium magnetic configuration (AFM). A $9 \times 9 \times 9$ grid of {\bf k}-points and an energy cutoff of $400$~eV has been used in the calculations.

To obtain a localized basis, we map our DFT Hamiltonian onto the basis of cubic harmonics represented by Wannier functions (WF). To this end, the \textsc{wannier90} code is employed.\cite{Mostofi2008} We use five $d$ orbitals for the TM atoms and three $p$ orbitals for oxygen. The resulting tight-binding Hamiltonian has the form,
\begin{equation}
H^{\sigma} = \sum_{i} \varepsilon_{i}^{\sigma}n_{i}^{\sigma} + \sum_{i \neq j} t_{ij}^{\sigma} c_{i}^{\sigma \dagger} c_{j}^{\sigma},
\label{eqn:TBHamiltonian}
\end{equation}
where $\sigma$ labels the spin projection, $\varepsilon_{i}^{\sigma}$ is the energy of the $i^{\rm th}$ WF, and $n_{i}^{\sigma}$ is its occupation number. $t_{ij}^{\sigma}$ is the hopping parameter between the $i^{\rm th}$ and $j^{\rm th}$ WF and $c_{i}^{\sigma \dagger}$ ($c_{j}^{\sigma}$) is the creation (annihilation) operator of electrons localized on the $i^{\rm th}$ ($j^{\rm th}$) WF. 
The local magnetic moments $M_{i}$ are calculated from the DFT density of states $g_{i}^{\sigma}(\varepsilon)$ projected onto the $i^{\rm th}$ WF, as 
\begin{equation}
M_{i} = \int_{-\infty}^{E_{F}} d\varepsilon \left[ g_i^{\uparrow}(\varepsilon) - g_i^{\downarrow}(\varepsilon) \right],
\label{eqn:MagneticMoment}
\end{equation}
where $E_F$ is the Fermi energy.

\subsection{Magnetic force theorem}
We consider the mapping onto the classical Heisenberg Hamiltonian in the limit of small angles: 
\begin{equation}
H = - \sum_{i > j} 2J_{ij} \bi{s}_i \cdot \bi{s}_j,
\label{eqn:HeisenbergHamiltonian}
\end{equation}
where $\bi{s}_i$ ($\bi{s}_j$) is the unit vector in the direction of the magnetic moment on site $i$ ($j$). $J_{ij}$ is the corresponding exchange interaction between sites $i$ and $j$. 

In the MFT method, the exchange interactions can be written in the following form:\cite{Lichtenstein1987}
\begin{equation}
J_{ij} = \frac{1}{4\pi} \int_{-\infty}^{E_{F}} d \varepsilon \sum_{m,m' \atop m'',m'''} {\rm Im} \left[ \Delta_{i}^{mm'} G_{ij \downarrow}^{m' m''} (\varepsilon) \Delta_{j}^{m'' m'''} G_{ji \uparrow}^{m''' m } (\varepsilon) \right],
\label{eqn:Exchange}
\end{equation}
where $\Delta_{i}^{mm'} = \int_{BZ} [H_{ii,\uparrow}^{mm'}(\bi{k})-H_{ii,\downarrow}^{mm'}(\bi{k}) ]d\bi{k}$ is the exchange splitting and $G_{ij \downarrow}^{mm'} (\varepsilon)$ is the real-space Green's function, that is calculated in reciprocal space by: 
\begin{equation}
G_{\mathbf{k}\sigma}(\varepsilon) = \left[ \varepsilon - H_{\sigma}(\bi{k}) +i\eta \right]^{-1},
\label{eqn:GreenFunctionK}
\end{equation}
where $\eta = 1$ meV is a smearing parameter and $H_{\sigma}(\bi{k})$ is the reciprocal Hamiltonian matrix defined in orbital space whose elements are obtained from the DFT calculations. For the calculations in reciprocal space, a $6 \times 6 \times 6$ {\bf k}-mesh has been used, which is sufficient to obtain the converged values of $J_{ij}$. The calculations of exchange integrals within the MFT method are done with an in-house developed code.\cite{Rudenko2013}

To estimate the N\'eel temperature $T_N$ in hematite, we considered the mean field approximation and calculated $T_N$ in accordance with \cite{Anderson1963} as the largest eigenvalue of the following matrix,
\begin{equation}
\theta_{ij} = \frac{J_{ij}s_{i}s_{j}}{3k_{B}},
\label{Eq:NeelTemperature}
\end{equation}
where $k_B$ is the Boltzmann constant.

\subsection{Total energy calculations}
\begin{figure}[hb]
\centering
\includegraphics[height=5cm]{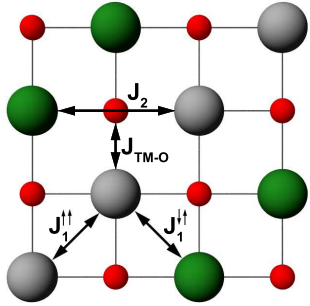}%
\caption{\label{fig:ExchangeRockSalt}(Color online) A 100 plane of the rock salt structure of NiO and MnO. The gray (green) atoms indicate the TM sites with magnetic moments up (down) respectively. Oxygen is shown in red. The exchange interactions are indicated by black arrows. }
\end{figure}

In addition to the MFT, we also used the total energy approach to calculate the exchange interactions. In this approach, we calculate the energies of multiple magnetic configurations using DFT and fit them to the Heisenberg model. For NiO and MnO, we used three magnetic states: (i) FM; (ii) AFI: AFM order in the [001] direction; and (iii) AFII: AFM order in the [111] direction, being the magnetic ground state. The corresponding total energies can be expressed as follows,
\begin{equation}
\eqalign{E_{\rm FM} = E_{0} -12J_1 -6J_2 \cr
E_{\rm AFI} = E_{0} +4J_1 -6J_2 \cr
E_{\rm AFII} = E_{0} +6J_2.}
\end{equation}
where $J_1$ and $J_2$ can be identified using figure~\ref{fig:ExchangeRockSalt} and $E_0$ corresponds to the non-magnetic part of the energy. Using these energies, the exchange interactions per TM-oxygen pair can be calculated as
\begin{equation}
\eqalign{J_1 = \frac{1}{16}\left(E_{\rm AFI}-E_{\rm FM}\right) \cr
J_2 = \frac{1}{48}\left(4E_{\rm AFII}-E_{\rm FM}-3E_{\rm AFI}\right).}
\end{equation}

\begin{figure}[ht]
\centering
\includegraphics[height=6cm]{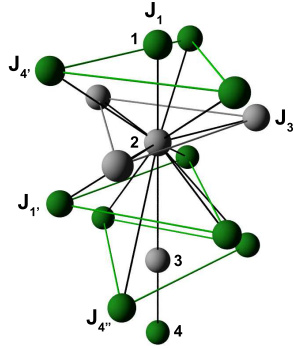}%
\caption{\label{fig:ExchangeHematite}(Color online) Fe atoms within a single unit cell of hematite (central line with numbers $1-4$), where gray (green) indicates magnetic moments up (down), respectively. The exchange interactions of Fe$_2$ with other Fe atoms are also shown where the single (double) apostrophe indicates the (next) nearest neighbor unit cell and the number the corresponding Fe atom in the unit cell.}
\end{figure}

In hematite, we only consider five different magnetic configurations within the unit cell: $+--+$ (ground state), $--++$, $+-+-$, $-+++$ and $++++$ where the plus and minus indicate the spin directions of the four Fe atoms in the unit cell, ordered as shown in figure~\ref{fig:ExchangeHematite}. The total energy can be expressed as follows:
\begin{eqnarray}
E_{++++} = E_{0} -J_1 -J_3 -3(J_{1'} +J_{3'} +J_{4'} +J_{4''}) \label{Eq:TotEn:HematiteEnergy}\nonumber\\
E_{+--+} = E_{0} +J_1 -J_3 +3(J_{1'} -J_{3'} +J_{4'} +J_{4''})\nonumber\\
E_{--++} = E_{0} -J_1 +J_3 -3(J_{1'} -J_{3'} -J_{4'} -J_{4''})\\
E_{+-+-} = E_{0} +J_1 +J_3 +3(J_{1'} +J_{3'} -J_{4'} -J_{4''})\nonumber\\
E_{-+++} = E_{0} +J_1 -J_3 +3(J_{1'} -J_{3'} -J_{4'} -J_{4''}),\nonumber
\end{eqnarray}
where $J_1$, $J_3$, $J_{1'}$, $J_{3'}$, $J_{4'}$ and $J_{4''}$ can be identified using figure~\ref{fig:ExchangeHematite}.
Note, we do not distinguish between the exchange interactions in different unit cells, and we only resolve the following interactions: 
\begin{eqnarray}
J_{1}^{\rm TE} &= J_{1} + 3J_{1'},\nonumber\\
J_{3}^{\rm TE} &= J_{3} + 3J_{3'},\label{eq:exchange_hematite_toten}\\
J_{4}^{\rm TE} &= J_{4'} + J_{4''}. \nonumber
\end{eqnarray}
In principle, next nearest neighbor interactions could also be taken into account. However, this requires an enlargement of the unit cell, which would lead to at least 35 different AFM configurations with eight degrees of freedom. Within the single unit cell, we have four degrees of freedom ($E_0$, $J_{1}^{\rm TE}$, $J_{3}^{\rm TE}$ and $J_{4}^{\rm TE}$) and five equations. Therefore, we calculate the average and standard deviation for the exchange interactions for all possible combinations.

\subsection{Effective exchange interactions}
In the presence of spin polarization on oxygen, also TM-O exchange interactions occur. If no O-O interactions are present, the O sites follow the magnetic field created by the TM atoms and we can calculate the corresponding effective exchange interactions. In the Appendix it is shown that this leads to the following effective exchange interactions: 
\begin{equation}
J_{ij}^{\rm eff} = J_{ij} + \frac{2\sum_k J'_{ik}J'_{kj}}{|\sum_l J'_{il}|},
\label{ExchangeEff}
\end{equation}
where $i$ and $j$ ($k$ and $l$) label TM (O) sites, respectively. $J'_{ik}$ is the TM-O exchange interaction between TM site $i$ and O site $k$. In the rock salt structure, (\ref{ExchangeEff}) results in the following effective exchange interactions: 
\begin{equation}
\eqalign{J_1^{\rm eff} = J_1 + \frac{2}{3}J_{\rm TM-O}\nonumber \cr
J_2^{\rm eff} = J_2 + \frac{1}{3}J_{\rm TM-O}}
\label{Eq:effectiveExchangeRocksalt}
\end{equation}
In hematite, the effective exchange interactions are given by:
\begin{equation}
\eqalign{J_{1'}^{\rm eff} &= J_{1'} + \frac{2J_{{\rm Fe-O}_1}^2}{3(J_{{\rm Fe-O}_1}+J_{{\rm Fe-O}_2})}, \cr
J_{4'/4''}^{\rm eff} &= J_{4'/4''} + \frac{2J_{{\rm Fe-O}_1}J_{{\rm Fe-O}_2}}{3(J_{{\rm Fe-O}_1}+J_{{\rm Fe-O}_2})}, \cr
J_{1}^{\rm eff} &= J_{1} + \frac{6J_{{\rm Fe-O}_2}^2}{3(J_{{\rm Fe-O}_1}+J_{{\rm Fe-O}_2})}, \cr
J_{3'}^{\rm eff} &= J_{3'} + \frac{4J_{{\rm Fe-O}_2}^2}{3(J_{{\rm Fe-O}_1}+J_{{\rm Fe-O}_2})}, } 
\label{Eq:effectiveExchangeHematite}
\end{equation}
where $J_{{\rm Fe-O}_1}$ and $J_{{\rm Fe-O}_2}$ correspond to the two exchange interactions between Fe and O. 

\section{\label{sec:resultsdiscussion}Results}

\subsection{NiO}
The magnetic ground state of NiO is formed by AFM ordering along the [111] crystallographic axis. Using (\ref{eqn:MagneticMoment}), we calculate the magnetic moments in the AFM ground state and in the FM configuration for both the Ni and O sites. The Ni magnetic moments are 1.63~$\mu_B$ and 1.69~$\mu_B$ in the AFM and FM configuration, respectively. Whereas the Ni magnetic moments are independent on the magnetic state, the O sites show a pronounced dependence with 0.00~$\mu_B$ and 0.30~$\mu_B$ for the AFM and FM states, respectively. 
The calculated magnetic moments of the Ni sites agree very well with previous calculations using the local self- interaction correction (LSIC)\cite{Fischer2009}: 1.68~$\mu_{B}$ and LDA+DMFT\cite{Kvashnin2015}:  1.85~$\mu_{B}$., The experimental magnetic moment of Ni is larger: 1.90~$\mu_{B}$, but it comprises both spin and orbital contributions to the magnetization.\cite{Cheetham1983}

\begin{table}[hbpt]
  \caption{\label{tab:ExchangeNiO} Exchange interactions (in meV) in NiO calculated for the AFM and FM magnetic states using the MFT and total energy (Tot. En.) methods.}
    \begin{indented}
    \lineup
    \item[]\begin{tabular}{@{}llll}
    \br
    & \centre{2}{MFT} & Tot. En. \cr
    \cline{2-3}
    & AFM & FM & \cr
    \mr
    ${J}_{1}$   & \m$0.1$ & \m\0$0.0$ & \m\0$1.1$ \cr
    ${J}_{2}$   &  $-9.8$ & $-14.7$ & $-11.9$\cr
    ${J}_{\rm Ni-O}$  & \m$0.0$ & \m\0$5.3$ & \cr
    \mr
    ${J}_{1}^{\rm eff}$ & \m$0.1$ & \m\0$3.5$ & \cr
    ${J}_{2}^{\rm eff}$ & $-9.8$ & $-12.9$ & \cr
    \br
    \end{tabular}
    \end{indented}
\end{table}

Table~\ref{tab:ExchangeNiO} shows the exchange interactions for both the FM and AFM configurations calculated using the MFT as defined in (\ref{eqn:Exchange}).
In principle the nearest neighbor interaction $J_1$ can be spin orientation dependent as is indicated with ${J}_{1}^{\uparrow\downarrow}$ and ${J}_{1}^{\uparrow\uparrow}$ in figure~\ref{fig:ExchangeRockSalt}. However for NiO, ${J}_{1}^{\uparrow\downarrow}$ and ${J}_{1}^{\uparrow\uparrow}$ are equal and $J_{1} \equiv {J}_{1}^{\uparrow\downarrow} = {J}_{1}^{\uparrow\uparrow}$ is used in table~\ref{tab:ExchangeNiO}. 
In the AFM configuration, $J_{1}$ is weak and prefers FM alignment ($0.1$~meV), whereas $J_2$ is two orders of magnitude larger and is of the opposite sign ($-9.8$~meV). This strong AFM interaction is due to the overlap between the $3z^2-r^2$ orbitals on Ni and an intermediate $p_z$ orbital on O.\cite{Korotin2014} This is consistent with the Goodenough-Kanamori rules, which predict a small but FM interaction for the Ni-O-Ni bonds forming an angle of $90^\circ$ ($J_1$) and stronger AFM interactions between the linear $180^\circ$ Ni-O-Ni bonding ($J_2$).\cite{Goodenough1963}

In the FM configuration, when spin polarization on oxygen is neglected, $J_1$ reduces to $0.0$~meV and $J_2 = -14.7$~meV favors AFM ordering, even stronger compared to the AFM case. Therefore, even starting from the FM configuration, we obtain that AFM ordering is energetically favorable.
If the spin polarization on oxygen is taken into account, the absence of $J_{\rm Ni-O}$ interactions in the AFM case result in no renormalization of the Ni-Ni interactions. However, in the FM case, a non-negligible Ni-O interaction ($J_{\rm Ni-O} = 5.3$~meV) is obtained. Mapping this Ni-O exchange interaction onto an effective Ni-Ni model using (\ref{Eq:effectiveExchangeRocksalt}), we determine the effective exchange interactions between the Ni sites, shown in table~\ref{tab:ExchangeNiO}. $J_{2}^{\rm eff} = -12.9$~meV remains the dominant exchange interaction and favors AFM alignment of spins, whereas $J_{1}^{\rm eff} = 3.5$~meV becomes larger in magnitude compared to the AFM case. 

If we approximate the exchange interactions by the total energy method using the energies of the AFI, AFII and FM configurations, the exchange interactions can be determined as $J_1 = 1.1$~meV and $J_2 = -11.9$~meV. Comparing the MFT results and those of the total energy method, we find the total energy exchange interactions are inbetween the interactions obtained for the different configurations using the MFT.

NiO and MnO are well known for the strongly correlated nature of the $3d$ electrons in these materials. The simple rock-salt structure along with the inability of regular DFT functionals to correctly describe these materials, make NiO and MnO typical benchmark systems to test methods for the calculation of exchange interactions in strongly correlated systems. 
Beyond the DFT methods such as LDA$+U$,\cite{Zhang2006,Korotin2014,Jacobsson2013} hybrid functionals,\cite{Moreira2002,Archer2011} the self-interaction correction,\cite{Kodderitzsch2002a,Fischer2009} $GW$ approximation\cite{Faleev2004} and dynamical mean-field theory\cite{Kunes2007,Kvashnin2015} have been used successfully to improve the correspondence between calculations and experiments in these materials. 

\begin{table}[htbp]
  \caption{\label{tab:ExchangeNiORef} Exchange interactions (in meV) for NiO determined from experiment (Exp) and other calculation schemes. } 
    \begin{indented}
    \lineup
    \item[]\begin{tabular}{@{}lll}
    \br
    Method & $J_{1}$ & $J_{2}$ \\
    \mr
    Exp.\cite{Shanker1973,Hutchings1972}  & $-0.69,0.69$ & $-8.66,-9.51$ \\
    PBE$+U$\cite{Zhang2006} &  \m$0.87$ & $-9.54$ \\
    PBE$+U$\cite{Korotin2014} &  \m$0.20$ & $-9.45$ \\
    Fock35\cite{Moreira2002} & \m$0.95$ & $-9.35$ \\
    LSIC\cite{Fischer2009} & \m$0.15$ & $-6.92$ \\
    LSDA+DMFT\cite{Kvashnin2015} & $-0.04$ & $-6.53$ \\
    \br
    \end{tabular}
    \end{indented}
\end{table}

Table~\ref{tab:ExchangeNiORef} shows exchange interactions obtained for the AFM ground state by other works for NiO. \cite{Zhang2006,Korotin2014,Moreira2002,Fischer2009,Kvashnin2015,Shanker1973,Hutchings1972}
It is well established that regular LDA is insufficient to calculate exchange interactions in NiO.\cite{Kvashnin2015,Archer2011,Zhang2006,Moreira2002} In contrast, LSIC overestimates the electron localization, leading to a slight underestimation of the exchange interactions. In \cite{Moreira2002} a comprehensive study on the effect of Fock exchange on the NiO parameters is performed and the best overall agreement is found for $35\%$ Fock exchange. 
In \cite{Zhang2006}, the PBE$+U$ functional ($U=6.3$~eV) is used to calculate the exchange interactions with the total energy method. In \cite{Korotin2014} the PBE$+U$ functional is also used but with a larger $U = 8$~eV and the MFT method to calculate the exchange interactions. Despite the differences in methods and $U$ values, all PBE$+U$ results lead to similar interaction strengths of $J_2$ in NiO and are close to the experimentally determined values of $-8.66$ and $-9.51$~meV. 
Surprisingly, dynamical electron correlations using LSDA+DMFT lead only to small corrections in the exchange interactions compared to LSDA$+U$ ($J_1 = -0.03$~meV and $J_2 = -6.8$~meV) for NiO and static local correlations capture most of the essential modifications.\cite{Kvashnin2015}

Among the different methods, the $J_1$ interaction shows a spread of $0.7$~meV. However, from the experimental point of view, the magnitude and even sign of $J_1$ remain uncertain. 
Overall, our results for $J_1$ and $J_2$ for the AFM configuration are very close to the experimental values and other works mentioned in table~\ref{tab:ExchangeNiORef}. The effective interactions for the FM configuration and the total energy interactions deviate more due to the consideration of different magnetic configurations. 

\subsection{MnO}
Like NiO, MnO crystallizes in the rock salt structure and has an AFM ordering along the [111] direction. In the AFM ground state, the magnetic moment of the Mn sites amounts to $4.69$~$\mu_{B}$ and no spin polarization on the O atom is observed. In the FM state, the magnetic moments are $4.72$ and $0.28$~$\mu_{B}$ for Mn and O, respectively. 
The experimental magnetic moment of Mn in MnO is slightly lower: 4.54~$\mu_B$. \cite{Jauch2004} The calculated magnetic moments for Mn are close to other methods: 4.63~$\mu_{B}$ (LSIC) and $4.52$~$\mu_{B}$ (PBE0). \cite{Fischer2009,Franchini2005} 

\begin{table}[htbp]
  \caption{\label{tab:ExchangeMnO} Calculated exchange interactions (in meV) in MnO for the AFM and FM magnetic state using the MFT and total energy methods.}
    \begin{indented}
    \lineup
    \item[]\begin{tabular}{@{}llll}
    \br
    & \centre{2}{MFT} & Tot. En.\\
    \cline{2-3}
    Atom & AFM & FM & \\
    \mr
    ${J}_{1}^{\uparrow\uparrow}$   & $-2.6$ & $-2.8$ & \multirow{2}{*}{$-2.8$ }\\
    ${J}_{1}^{\uparrow\downarrow}$ & $-3.1$ & $-3.0$ & \\
    ${J}_{2}$   & $-2.6$ & $-3.0$ & $-1.4$\\
    ${J}_{\rm Mn-O}$  &  \m$0.0$ & \m$3.9$&  \\
    \mr
    ${J}_{1}^{\rm eff}(\uparrow\uparrow)$ & $-2.6$ & $-0.2$ & \\
    ${J}_{1}^{\rm eff}(\uparrow\downarrow)$ & $-3.1$ & $-0.4$ & \\
    ${J}_{2}^{\rm eff}$ & $-2.6$ & $-1.7$ & \\
    \br
    \end{tabular}
    \end{indented}
\end{table}

Table~\ref{tab:ExchangeMnO} shows the calculated exchange interactions for MnO. In the AFM ground state, we observe two different $J_{1}$ exchange interactions: between parallel spins in the same fcc plane $J_{1}^{\uparrow\uparrow} = -2.6$~meV and between antiparallel spins in neighboring fcc planes $J_{1}^{\uparrow\downarrow} = -3.1$~meV. For visual aid $J_{1}^{\uparrow\uparrow}$ and $J_{1}^{\uparrow\downarrow}$ are also indicated in figure~\ref{fig:ExchangeRockSalt}.  Next nearest neighbor exchange, $J_2 = -2.6$~meV is of the same order as $J_{1}^{\uparrow\uparrow}$ and $J_{1}^{\uparrow\downarrow}$. 
The difference between $J_{1}^{\uparrow\uparrow}$ and $J_{1}^{\uparrow\downarrow}$ has been observed before in both calculations\cite{Franchini2005,Jacobsson2013} and experiments\cite{Pepy1974} and has been attributed to the distorted trigonal structure. In our method, however, the difference between $J_{1}^{\uparrow\uparrow}$ and $J_{1}^{\uparrow\downarrow}$ is not the result of a geometric distortion (we used an undistorted structure), but is purely an electronic effect. Resolving the orbital contributions of the exchange interactions for the AFM configuration, we obtain that $J_{1}^{\uparrow\uparrow}$ consists of only $t_{2g}-t_{2g}$ interactions, whereas in $J_{1}^{\uparrow\downarrow}$ two contributions are present: $-2.8$~meV being the result of $t_{2g}-t_{2g}$ interactions, and $-0.4$~meV, which is due to $e_{g}-e_{g}$ interactions. On the contrary, in $J_2$, $e_{g}-e_{g}$ interactions constitute the majority of the interaction ($-2.1$~meV) and only a small contribution ($-0.4$~meV) is due to $t_{2g}-t_{2g}$. 

In the FM configuration, both $J_1^{\uparrow\uparrow}$ and $J_2$ turn out to be slightly larger, corresponding to $-2.8$ and $-3.0$~meV, respectively. As expected, the additional interaction $J_{\rm Mn-O} = 3.9$~meV is also found. 
$J_1^{\uparrow\uparrow}$ and $J_1^{\uparrow\downarrow}$ consist predominantly of $t_{2g}-t_{2g}$ interactions, amounting to $-3.0$ and $-2.8$~meV, respectively. Both $J_1^{\uparrow\uparrow}$ and $J_1^{\uparrow\downarrow}$ contain also small $e_{g}-e_{g}$ ($-0.1$~meV) and $t_{2g}-e_{g}$ ($0.1$~meV) contributions.
The majority of the $J_2$ interaction is formed by $e_{g}-e_{g}$ ($-2.8$~meV) interactions and a small fraction of $t_{2g}-t_{2g}$ ($-0.2$~meV). The $J_{\rm Mn-O}$ interaction is comprised of $1.7$ and $2.3$~meV contributions, corresponding to $t_{2g}-p$ and $e_{g}-p$, respectively. 

If oxygen spin polarization is taken into account, the positive $J_{\rm Mn-O}$ interaction reduces the effective exchange interactions as defined by (\ref{Eq:effectiveExchangeRocksalt}) to $J_{1}^{\rm eff}(\uparrow\uparrow) = -0.2$~meV, $J_{1}^{\rm eff}(\uparrow\downarrow) = -0.4$~meV, and $J_{2}^{\rm eff} = -1.7$~meV. Therefore, spin polarization on oxygen leads to considerable changes in the effective Mn-Mn exchange interactions. The effect is most pronounced ($\sim$2.5~eV) for the nearest neighbor interactions $J_{1}^{\rm eff}(\uparrow\uparrow)$ and $J_{1}^{\rm eff}(\uparrow\downarrow)$, which are reduced by an order of magnitude. For $J_{2}^{\rm eff}$, a change of $53\%$ is observed. 

\begin{table}[htbp]
  \caption{\label{tab:ExchangeMnORef} Exchange interactions (in meV) for MnO determined from experiment (Exp) and other calculation schemes. } 
  \begin{indented}
  \lineup
    \item[]\begin{tabular}{@{}llllll}
    \br
     & Exp\cite{Pepy1974} & PBE$+U$\cite{Franchini2005} & PBE0\cite{Franchini2005} & LSIC\cite{Fischer2009} & LDA$+U$\cite{Jacobsson2013} \\
    \mr
    ${J}_{1}$                     &       & $-2.2$ & $-3.1$ & $-0.9$ & $-2.9$ \\
    ${J}_{1}^{\uparrow\uparrow}$  & $-2.01$ & $-2.1$ & $-2.9$ &  & $-2.6$ \\
    ${J}_{1}^{\uparrow\downarrow}$& $-2.65$ & $-2.3$ & $-3.5$ &  & \m$3.1$ \\
    ${J}_{2}$                     & $-2.79$ & $-1.2$ & $-3.7$ & $-2.0$ & $-3.0$ \\
    \br
    \end{tabular}
    \end{indented}
\end{table}

Table~\ref{tab:ExchangeMnORef} shows the experimental results and that of various calculation methods for MnO. If we compare our AFM results to the experimentally fitted values, $J_2$ is very similar. However, the magnitude of both ${J}_{1}^{\uparrow\uparrow}$ and ${J}_{1}^{\uparrow\downarrow}$ is slightly overestimated by $\sim$0.5~meV in our calculations. Note that PBE$+U$ using the total energy method gives the wrong order of $J_1$ and $J_2$, both in our calculations and in \cite{Franchini2005}. PBE0 using the total energy method gives the correct order, but overestimates the strength both for $J_1$ and $J_2$. In contrast, the LSIC calculations in \cite{Fischer2009} underestimate both exchange interactions. LDA$+U$ using spin spiral calculations give similar results for $J_1$, but our results underestimate $J_2$, whereas LSIC leads to a small overestimation. 

\subsection{Fe$_2$O$_3$}
Hematite crystallizes into a trigonal crystal structure. At $T < 250$~K the magnetic moments of Fe atoms align along the trigonal axis, forming an AFM ordering (see  figure~\ref{fig:ExchangeHematite}).
We calculate the magnetic moment in this AFM configuration to be $4.09$~$\mu_{B}$ for Fe and observe no spin polarization on O. 
When we consider the FM configuration, the magnetic moments of Fe increase to $4.20$~$\mu_B$ and a spin polarization of $0.53$~$\mu_B$ is observed on the O sites. 
The magnetic moments of 4.09~$\mu_{B}$ (AFM) and 4.20~$\mu_{B}$ (FM) are in excellent agreement with previous works for both the AFM state (4.1~$\mu_{B})$\cite{Rollmann2004,Mazurenko2005} and the FM state (4.2~$\mu_{B}$).\cite{Rollmann2004} The experimental magnetic moments are slightly larger: 4.6--4.9~$\mu_{B},$\cite{Kren1965,Coey1971} which can, again, be explained by the orbital contribution.

\begin{table}[htbp]
  \caption{\label{tab:ExchangeHematite}Exchange parameters (in meV) in hematite for the AFM and FM magnetic configurations as defined in figure~\ref{fig:ExchangeHematite}. $n_i$ and $d_i$ are the coordination number and interatomic distance for a given interaction $J_i$, respectively. }
    \begin{indented}
    \lineup
    \item[]\begin{tabular}{@{}lllll}
    \br
     & AFM & FM & $n_i$ & $d_{i}$ (\AA) \\
    \mr
    $J_{1'}$ & $-13.9$ & $-18.8$  & 3 & $3.41$\\
    $J_{4'} \& J_{4''}$  &  \0$-9.8$ & $-14.5$ & 6 & $3.73$\\
    $J_{1}$ &  \0$-3.5$ & \0$-3.0$ & 1 & $2.89$\\
    $J_{3'}$ & \0$-3.2$ & \0$-4.2$ & 3 & $2.98$\\
    $J_{{\rm Fe}-{\rm O}_1}$ &   \m\0$0.0$ &  \m$20.9$ & 3 & $1.97$\\
    $J_{{\rm Fe}-{\rm O}_2}$ &   \m\0$0.0$ &  \m$13.6$ & 3 & $2.12$\\
    \br
    \end{tabular}
    \end{indented}
\end{table}

In hematite, five relevant Fe-Fe exchange interactions between Fe sites are present and can be identified using figure~\ref{fig:ExchangeHematite}. The Fe atoms in the unit cell on the vertical axis are denoted by numbers $1$--$4$. The exchange interactions between atom 2 and other Fe atoms are shown in figure~\ref{fig:ExchangeHematite}, where a single (double) apostrophe indicates the (next) nearest neighbor unit cell, and the number corresponds to the specific Fe atom in the unit cell forming a pair with atom 2. Table~\ref{tab:ExchangeHematite} shows the exchange interactions in hematite calculated for both the AFM and FM configurations. In the AFM case, the largest exchange interaction ${J}_{1'} = -13.9$~meV corresponds to the interatomic distance $d_1=3.41$~\AA. The second largest exchange interaction corresponds to the interaction between atom 2 and atom 4 in the nearest ($J_{4'}$) and next nearest ($J_{4''}$) unit cell and it amounts to $-9.8$~meV. Interaction between atom 2 and atom 1, $J_{1} = -3.5$~meV happens to be the only relevant exchange interaction within the unit cell. Indeed, $J_3$ is very small due to the large ($\sim$4.1~\AA) interatomic distance $d_3$ between atoms 2 and 3. In contrast, the interaction with atomc 3 from the adjacent cell is larger with $J_{3'} = -3.2$~meV. In the AFM configuration no spin polarization on oxygen is present and, therefore, $J_{{\rm Fe}-{\rm O}_1}$ and $J_{{\rm Fe}-{\rm O}_2}$ are $0.0$~meV. 

In the FM configuration, ${J}_{1'} = -18.8$~meV favors even stronger AFM coupling than in the AFM case. The same is true for $J_{4'}$ and $J_{4''}$ with $-14.5$~meV compared to $-9.8$~meV in the AFM case. Smaller interactions $J_{1}$ and $J_{3'}$ are only slightly different, yielding $-3.0$~meV and $-4.2$~meV, respectively. In the presence of spin polarization on oxygen, two different Fe-O exchange interactions: $20.9$ and $13.6$~meV are obtained, corresponding to the interatomic distances $1.97$ and $2.12$~\AA{}, respectively. It is worth noting that these exchange interactions between Fe and O are much stronger compared to NiO and MnO. 
The corresponding effective exchange interactions are determined using \ref{Eq:effectiveExchangeHematite} and shown in table~\ref{tab:ExchangeHematiteEff}. For $J_{1'}^{\rm eff}$, $J_{4'}^{\rm eff}$, and $J_{4''}^{\rm eff}$, the relative change between AFM and FM configurations is $34\%$, $9\%$ and $9\%$, respectively. On the contrary, for $J_{1}^{\rm eff}$ and $J_{3'}^{\rm eff}$, the opposite sign and large difference ($-144\%$ and $-146\%$) is obtained. 

\begin{table}[htbp]
  \caption{\label{tab:ExchangeHematiteEff}Effective exchange parameters (in meV) in hematite calculated for the AFM and FM magnetic configurations, and their relative difference ($=({\rm AFM}-{\rm FM})/{\rm FM}$). }
    \begin{indented}
    \lineup
    \item[]\begin{tabular}{@{}llll}
    \br
     & AFM & FM & Difference ($\%$)\\
    \mr
    $J_{1'}^{\rm eff}$ & $-13.9$ & $-10.4$ & \m\0$34$ \\
    $J_{4'}^{\rm eff} \& J_{4''}^{\rm eff}$  &  \0$-9.8$ & \0$-9.0$ & \m\0\0$9$\\
    $J_{1}^{\rm eff}$ &  \0$-3.5$ &  \m\0$7.8$ & $-144$\\
    $J_{3'}^{\rm eff}$ &  \0$-3.1$ &  \m\0$6.8$ & $-146$\\
    \br
    \end{tabular}
    \end{indented}
\end{table}

Experimental exchange interactions derived from inelastic neutron scattering\cite{Samuelsen1970} and previously calculated exchange interactions using LDA$+U$\cite{Mazurenko2005} are summarized in table~\ref{tab:ExchangeHematiteRef}. Whereas the exchange interactions in \cite{Mazurenko2005} are larger than the experimental fitted values, our results for the AFM configuration are slightly lower. For the four principal exchange interactions: $J_{1'}$, $J_{4'}$, $J_{4''}$, and $J_{1}$ our results are closer to the experimentally derived values. Our estimation of the N\'eel temperature $T_N=878$~K calculated in the mean field approximation [see (\ref{Eq:NeelTemperature})] is in excellent agreement with experiments (947--969~K). The underestimation of the calculated N\'eel temperature by $9\%$ indicates too low exchange interactions, whereas the mean field approximation usually overestimates the N\'eel temperature.  The N\'eel temperature obtained in \cite{Mazurenko2005} is overestimated, which is supposedly attributed to a different approach used in \cite{Mazurenko2005} for the calculation of both exchange integrals and the N\'eel temperature.

\begin{table}[h]
  \caption{\label{tab:ExchangeHematiteRef}Experimentally determined\cite{Samuelsen1970} and previously calculated\cite{Mazurenko2005} exchange interactions using LDA$+U$ for hematite (in meV).}
    \begin{indented}
    \lineup
    \item[]\begin{tabular}{@{}lll}
    \br
     & Exp\cite{Samuelsen1970} & LDA+$U$\cite{Mazurenko2005} \\
    \mr
    $J_{1'}$ & $-17.1$ & $-25.2$ \\
    $J_{4'} \& J_{4''}$  &  $-12.6$ & $-17.5$ \\
    $J_{1}$ &  \m\0$3.1$ &  \0$-8.6$ \\
    $J_{3}$ &  \0$-1.1$ &  \m\0$0.1$ \\
    $J_{3'}$ &  \m\0$0.52$ &  \m\0$7.3$ \\
    \br
    \end{tabular}
    \end{indented}
\end{table}

Finally, we also calculated the exchange interactions in hematite using the total energy method as defined by (\ref{eq:exchange_hematite_toten}). For simplicity, we only consider different magnetic configurations within the unit cell, as described in section~\ref{sec:method}~C with energies given by (\ref{Eq:TotEn:HematiteEnergy}).
Table~\ref{tab:ExchangeHematiteTotEn} shows the corresponding exchange interactions. One can see that the correspondence between the total energy and MFT for the FM configuration is better than that with the MFT for the AFM configuration. This is due to the significant number of the ferrimagnetic and FM states considered in the total energy approach. The dependence on the magnetic states can also be seen in the extremely large standard deviation in the exchange interactions obtained. 

\begin{table}[t]
  \caption{\label{tab:ExchangeHematiteTotEn} The exchange interactions (in meV) calculated by the total energy method, as defined in (\ref{eq:exchange_hematite_toten}). For comparison, the results of MFT are also shown.}
    \begin{indented}
    \lineup
    \item[]\begin{tabular}{@{}lllll}
    \br
     & Tot. En. & MFT AFM & MFT FM\\
    \mr
    $J_{1}^{\rm TE}$ & $-29.1 \pm 65.9$ & $-45.2$ & $-23.4$ \\
    $J_{3}^{\rm TE}$ & \m$20.6 \pm 65.9$ & \0$-9.3$ & \m$20.4$\\
    $J_{4}^{\rm TE}$ & $-17.1 \pm 22.0$ &  $-19.6$ & $-18.0$\\
    \br
    \end{tabular}
    \end{indented}
\end{table}

\section{\label{sec:disc}Discussion}
The effect of oxygen spin polarization on magnetic states has been previously studied in the case of pyroxenes and LiCu$_2$O$_2$, where the magnetic moment on oxygen contributes to the stabilization of the FM configuration. In the AFM configuration, both spin up and down would hop from the O-$p$ level to the TM atoms, whereas in the FM case, only spin down can hop. As a result, the Hund's rule coupling on oxygen will gain more energy in the FM configuration.\cite{Mazurenko2007,Streltsov2008} 
To estimate this effect in NiO and MnO, we approximated the intraatomic exchange interaction of the oxygen atom $J_p^H$ through the shift of the oxygen $2p$ band centers for spin up $C_{\uparrow}$ and spin down $C_{\downarrow}$, which can be obtained from the DFT calculation. The resulting intraatomic exchange reads: $J_p^H = (C_{\uparrow}-C_{\downarrow})/M_{O}$, where $M_{O}$ is the spin polarization on oxygen.\cite{Streltsov2008} The corresponding energy $E_{O} = -IM_{O}^2/4$ is found to be $39$ and $10$~meV in NiO and MnO, respectively. The total energy exchange interactions without the Hund's rule coupling on oxygen become $J_1 = -1.3$~meV, $J_2 = -12.7$~meV for NiO and $J_1 = -3.5$~meV, $J_2 = -1.4$~meV for MnO if we subtract $E_{O}$ from the energy of the FM configuration. The influence of Hund's rule coupling on the oxygen atoms on the exchange interactions is, therefore, not essential.

In NiO and MnO, the spin polarization on oxygen is similar (0.3~$\mu_B$). Although the larger magnetic moment of Mn leads to somewhat stronger oxygen spin polarization, the increase in the lattice distance for MnO cancels the net result. In hematite the spin polarization is larger, $m_{{\rm O}} = 0.5$~$\mu_B$, due to the large magnetic moment on Fe and the smaller Fe-O distances ($1.97$--$2.12$~\AA) compared to MnO. 

Within the effective Heisenberg model derived in this work, which involves the renormalization of TM-TM exchange interactions in the presence of oxygen spin polarization, we observe an increase in the sensitivity of secondary (small in magnitude) exchange interactions on the magnetic state. These findings are in agreement with the recent results on CrO$_2$, where the largest exchange interaction only differs by $11\%$, and the smaller exchange interactions are more sensitive to the magnetic state considered.\cite{Solovyev2015} In NiO and MnO, the smallest relevant exchange parameter $J_{1}^{\rm eff}$ shows the largest dependence on the magnetic state, which can be explained as follows. First, the $J_{1}^{\rm eff}$ interaction between TM sites involves two intermediate O atoms, increasing the net effect of spin polarization, whereas for $J_{2}^{\rm eff}$ only one O atom is involved [see (\ref{Eq:effectiveExchangeRocksalt})]. Second, the effect of oxygen spin polarization [$J_{\rm TM-O}$ in (\ref{Eq:effectiveExchangeRocksalt})] is independent of the original interaction ($J_1$ or $J_2$), which for NiO results in a considerably stronger effect since $|J_{1}^{\rm eff}| \ll |J_{2}^{\rm eff}|$. 

MnO and NiO are often considered as textbook examples for the realization of the superexchange spin coupling mechanism, as their magnetic moments are well localized and their crystal structure allows for easy application of the Goodenough-Kanamori rules. Based on the results presented in section~\ref{sec:resultsdiscussion}, we can conclude that in NiO, the presence of oxygen spin polarization results in effective exchange interactions, which are more suitable for the Heisenberg model. Indeed, the largest exchange interaction $J_2$ changes from the AFM to FM configuration by $50\%$ and $32\%$ without and with spin polarization, respectively.  
In MnO, the opposite situation occurs. The exchange interactions obtained ignoring spin polarization on oxygen make the corresponding Heisenberg model more applicable in the sense that different magnetic configurations (AFM or FM) can be considered with a single set of parameters. The largest difference between the two sets of parameters is observed for the $J_2$ interaction, and it reaches $15\%$. In the presence of spin polarization in MnO, $J_{1}^{\rm eff}(\uparrow\uparrow)$ and $J_{1}^{\rm eff}(\uparrow\downarrow)$ almost vanish in the FM configuration, contributing as $-0.2$ and $-0.4$~meV, respectively. In this case, $J_{2}^{\rm eff}$ changes by $53\%$ between the AFM and FM configuration. 

In hematite, both methods result in essentially non-Heisenberg behavior. If spin polarization on oxygen is neglected, $J_{1'}$, $J_{4'}$ and $J_{4''}$ change by $35\%$, $48\%$ and $48\%$, respectively. If oxygen spin polarization is taken into account, changes in $J_{1'}^{\rm eff}$, $J_{4'}^{\rm eff}$, and $J_{4''}^{\rm eff}$ are smaller: $34\%$, $9\%$, and $9\%$, respectively. For $J_{1}^{\rm eff}$, and $J_{3'}^{\rm eff}$, the difference between the AFM and FM situations is more significant and even results in a different sign of the exchange interaction. This also explains the large spread in the results obtained using the total energy method. 

In general, no universal trend in applicability of the Heisenberg model between the two different approaches can be observed. The practical choice between the methods should mainly depend on the (i) assumptions made during the mapping procedure, e.g., regarding the relevance of magnetic states causing spin-polarization of a ligand; and (ii) on the final application purpose of the Hamiltonian, including relevant energy range, temperatures, and phenomena to be described. For example, if the fraction of magnetic states in the final application with ligand spin polarization is low, spin polarization can naturally be ignored. In contrast, for systems with ligand spin polarization in the ground state, the fraction of relevant magnetic states with spin polarization is much higher and spin polarization might play a considerable role in the magnetic properties. 

The observed findings are especially useful for comparison with nanostructures such as for example atomic clusters, where the ground state often involves significant spin polarization of oxygen.\cite{Logemann2015} Furthermore, the proposed approaches can also be used to test the validity of the Heisenberg model used in atomistic spin dynamics simulations, where spin polarization on exchange-mediated ligands is often neglected. 

\section{\label{sec:conclusion}Conclusions}
In this work, we studied the role of oxygen spin polarization in the magnetic properties of typical AFM TM oxides, and assessed the applicability of the Heisenberg model to those materials. Specifically, we calculated the magnetic moments and exchange interactions in different magnetic configurations for NiO, MnO, and hematite using the MFT. We considered both the conventional picture, where spin polarization on oxygen is ignored, and derived a model, where oxygen is effectively included in the exchange interactions between metal sites. We found that small exchange interactions and interactions between TM sites with multiple bridging O atoms are most sensitive to the spin polarization of oxygen, which appears if an FM configuration of magnetic moments is considered.

For NiO, we found that the dominant next nearest neighbor exchange interaction $J_2$ increases considerably in the FM configuration with respect to the AFM ground state when spin polarization is ignored. This difference is reduced if spin polarization on oxygen is included. In contrast, the absence of oxygen spin polarization in MnO results in exchange interactions compatible with the Heisenberg model. If spin polarization on oxygen in MnO is included, a significant reduction of the exchange interactions obtained in the FM configuration compared to the AFM case is observed, which worsens the applicability of the Heisenberg model.
In hematite, both methods result in non-Heisenberg behavior of magnetic moments. Particularly, exchange interactions show strong orientation dependence and they change significantly between different magnetic configurations, independently of the presence of spin polarization of oxygen.
Our findings show no universal trend in the applicability of the Heisenberg model with respect to the ligand spin polarization in TM oxides. They suggest, however, that the practical choice of a more suitable approach for the exchange interactions should be governed by the particular application of the Heisenberg model, or should be considered in conjunction with a beyond-Heisenberg spin model.

\ack
We are thankful to V.~V. Mazurenko and Y.~O. Kvashnin for stimulating discussions. 
The Nederlandse Organisatie voor Wetenschappelijk Onderzoek (NWO) and SURFsara are acknowledged for the usage of the LISA supercomputer and their support. The work is supported by European Research Council (ERC) Advanced Grant No.~338957 FEMTO/NANO.

\appendix
\setcounter{section}{0}
\section{Effective exchange interactions}
Let us start with the following Heisenberg Hamiltonian: 
\begin{equation}
H = - \sum_{ij}J_{ij}\vec{e}_i \cdot \vec{e}_j - 2\sum_{ik} J_{ik}'\vec{e}_i \cdot \vec{e}_k',
\end{equation}
where $i$ and $j$ label TM sites and $k$ labels O sites, respectively. All TM-O interactions are indicated with $J'$. If we assume no O-O exchange interactions, the magnetic moment of the O sites just follows the local magnetic field created by the surrounding TM atoms ($\vec{h}_k$):
\begin{equation}
\vec{e}_k' \parallel \vec{h}_k = 2\sum_j J_{kj}' \vec{e}_j.
\end{equation} 
If we now use the identity $\delta \vec{e} = \delta\vec{h}/|h|$, we get:
\begin{equation}
\delta \vec{e}'_k = \frac{2\sum_j J_{kj}'\delta \vec{e}_j}{2|\sum_l J_{kl}'|},
\end{equation}
which can be written as the effective exchange interaction between the TM sites $i$ and $j$: 
\begin{equation}
J_{ij}^{\rm eff} = J_{ij} + \frac{2 \sum_{k}J_{ik}'J_{kj}'}{|\sum_l J_{lj}'|}.
\end{equation}

\section*{References}
\bibliographystyle{iopart-num}
\bibliography{Citations}

\providecommand{\newblock}{}
\begin{thebibliography}{10}
\expandafter\ifx\csname url\endcsname\relax
  \def\url#1{{\tt #1}}\fi
\expandafter\ifx\csname urlprefix\endcsname\relax\def\urlprefix{URL }\fi
\providecommand{\eprint}[2][]{\url{#2}}

\bibitem{Goodenough1963}
Goodenough J~B 1963 {\em {Magnetism and the Chemical Bond}\/} (New York/London:
  Interscience Publishers)

\bibitem{gubanov2012magnetism}
Gubanov V~A, Liechtenstein A~I and Postnikov A~V 2012 {\em {Magnetism and the
  Electronic Structure of Crystals}\/} (Springer Berlin Heidelberg)

\bibitem{Khomskii_2014}
Khomskii D~I 2014 {\em {Transition Metal Compounds}\/} (Cambridge: Cambridge
  University Press)

\bibitem{Mott1974}
Mott N~F 1974 {\em {Metal-Insulator Transitions}\/} (London: Taylor {\&}
  Francis)

\bibitem{Imada1998}
Imada M, Fujimori A and Tokura Y 1998 {\em Rev. Mod. Phys.\/} {\bf 70}
  1039--1263

\bibitem{vonsovskii1974magnetism}
Vonsovskii S~V 1974 {\em {Magnetism}\/} (John Wiley)

\bibitem{Kei1996}
Kei Y 1996 {\em {Theory of Magnetism}\/} (Springer-Verlag Berlin Heidelberg)

\bibitem{White2007}
White R~M 2007 {\em {Quantum Theory of Magnetism}\/} (Springer-Verlag Berlin
  Heidelberg)

\bibitem{Liechtenstein1984}
Liechtenstein A~I, Katsnelson M~I and Gubanov V~A 1984 {\em J. Phys. F\/} {\bf
  14} L125--L128

\bibitem{Lichtenstein1987}
Lichtenstein A~I, Katsnelson M~I, Antropov V~P and Gubanov V~A 1987 {\em J.
  Magn. Magn. Mater.\/} {\bf 67} 65--74

\bibitem{Katsnelson2000}
Katsnelson M~I and Lichtenstein A~I 2000 {\em Phys. Rev. B\/} {\bf 61}
  8906--8912

\bibitem{Turek2001}
Pajda M, Kudrnovsk{\'{y}} J, Turek I, Drchal V and Bruno P 2001 {\em Phys. Rev.
  B\/} {\bf 64} 174402

\bibitem{Turek2006a}
Turek I, Kudrnovsk{\'{y}} J, Drchal V and Bruno P 2006 {\em Philos. Mag.\/}
  {\bf 86} 1713--1752

\bibitem{Katsnelson2008}
Katsnelson M~I, Irkhin V~Y, Chioncel L, Lichtenstein A~I and {De Groot} R~A
  2008 {\em Rev. Mod. Phys.\/} {\bf 80} 315--378

\bibitem{Sato2010}
Sato K, Bergqvist L, Kudrnovsk{\'{y}} J, Dederichs P~H, Eriksson O, Turek I,
  Sanyal B, Bouzerar G, Katayama-Yoshida H, Dinh V~A, Fukushima T, Kizaki H and
  Zeller R 2010 {\em Rev. Mod. Phys.\/} {\bf 82} 1633--1690

\bibitem{Ebert2011}
Ebert H, K{\"{o}}dderitzsch D and Min{\'{a}}r J 2011 {\em Rep. Prog. Phys.\/}
  {\bf 74} 096501

\bibitem{Turzhevskii1990}
Turzhevskii A, Lichtenstein A~I and Katsnelson M~I 1990 {\em Sov. Phys. Solid
  State\/} {\bf 32} 1138

\bibitem{Boukhvalov2004}
Boukhvalov D~W, Dobrovitski V~V, Katsnelson M~I, Lichtenstein A~I, Harmon B~N
  and K{\"{o}}gerler P 2004 {\em Phys. Rev. B\/} {\bf 70} 054417

\bibitem{Boukhvalov2002}
Boukhvalov D~W, Lichtenstein A~I, Dobrovitski V~V, Katsnelson M~I, Harmon B~N,
  Mazurenko V~V and Anisimov V~I 2002 {\em Phys. Rev. B\/} {\bf 65} 184435

\bibitem{Mazurenko2014}
Mazurenko V~V, Kvashnin Y~O, Jin F, {De Raedt} H~A, Lichtenstein A~I and
  Katsnelson M~I 2014 {\em Phys. Rev. B\/} {\bf 89} 214422

\bibitem{Fedorova2015}
Fedorova N~S, Ederer C, Spaldin N~A and Scaramucci A 2015 {\em Phys. Rev. B\/}
  {\bf 91} 165122

\bibitem{Coldea2001}
Coldea R, Hayden S~M, Aeppli G, Perring T~G, Frost C~D, Mason T~E, Cheong S~W
  and Fisk Z 2001 {\em Phys. Rev. Lett.\/} {\bf 86} 5377--5380

\bibitem{Wysocki2010}
Wysocki A~L, Belashchenko K~D and Antropov V~P 2010 {\em Nat. Phys.\/} {\bf 7}
  485

\bibitem{Skubic2008}
Skubic B, Hellsvik J, Nordstr{\"{o}}m L and Eriksson O 2008 {\em J. Phys.
  Condens. Matter\/} {\bf 20} 315203

\bibitem{Mikhaylovskiy2015}
Mikhaylovskiy R~V, Hendry E, Secchi A, Mentink J~H, Eckstein M, Wu A, Pisarev
  R~V, Kruglyak V~V, Katsnelson M~I, Rasing T and Kimel A~V 2015 {\em Nat.
  Commun.\/} {\bf 6} 8190

\bibitem{Streltsov2008}
Streltsov S~V and Khomskii D~I 2008 {\em Phys. Rev. B\/} {\bf 77} 064405

\bibitem{Solovyev2015}
Solovyev I~V, Kashin I~V and Mazurenko V~V 2015 {\em Phys. Rev. B\/} {\bf 92}
  144407

\bibitem{Mryasov2005}
Mryasov O~N 2005 {\em Phase Transitions\/} {\bf 78} 197--208

\bibitem{Anderson1959}
Anderson P~W 1959 {\em Phys. Rev.\/} {\bf 115} 2--13

\bibitem{Kresse1996}
Kresse G and Furthm{\"{u}}ller J 1996 {\em Phys. Rev. B\/} {\bf 54}
  11169--11186

\bibitem{Blochl1994}
Bl{\"{o}}chl P~E 1994 {\em Phys. Rev. B\/} {\bf 50} 17953--17979

\bibitem{Kresse1999}
Kresse G and Joubert D 1999 {\em Phys. Rev. B\/} {\bf 59} 1758

\bibitem{Dudarev1998}
Dudarev S~L, Botton G~A, Savrasov S~Y, Humphreys C~J and Sutton A~P 1998 {\em
  Phys. Rev. B\/} {\bf 57} 1505--1509

\bibitem{Zhang2006}
Zhang W~B, Hu Y~L, Han K~L and Tang B~Y 2006 {\em Phys. Rev. B\/} {\bf 74}
  054421

\bibitem{Franchini2005}
Franchini C, Bayer V, Podloucky R, Paier J and Kresse G 2005 {\em Phys. Rev.
  B\/} {\bf 72} 045132

\bibitem{Lichtenstein1995}
Lichtenstein A~I, Anisimov V~I and Zaanen J 1995 {\em Phys. Rev. B\/} {\bf 52}
  5467--5471

\bibitem{Blake1966}
Blake R~L, Hessevick R~E, Zoltai T and Finger L~W 1966 {\em Amer. Mineral.\/}
  {\bf 51} 123--129

\bibitem{Mostofi2008}
Mostofi A~A, Yates J~R, Lee Y~S, Souza I, Vanderbilt D and Marzari N 2008 {\em
  Comput. Phys. Commun.\/} {\bf 178} 685--699

\bibitem{Rudenko2013}
Rudenko A~N, Keil F~J, Katsnelson M~I and Lichtenstein A~I 2013 {\em Phys. Rev.
  B\/} {\bf 88} 081405(R)

\bibitem{Anderson1963}
Anderson P~W 1963 {\em Solid State Phys.\/} {\bf 14} 99--214

\bibitem{Fischer2009}
Fischer G, D{\"{a}}ne M, Ernst A, Bruno P, Lueders M, Szotek Z, Temmerman W and
  Hergert W 2009 {\em Phys. Rev. B\/} {\bf 80} 014408

\bibitem{Kvashnin2015}
Kvashnin Y~O, Gr{\aa}n{\"{a}}s O, {Di Marco} I, Katsnelson M~I, Lichtenstein
  A~I and Eriksson O 2015 {\em Phys. Rev. B\/} {\bf 91} 125133

\bibitem{Cheetham1983}
Cheetham A~K and Hope D~A~O 1983 {\em Phys. Rev. B\/} {\bf 27} 6964--6967

\bibitem{Korotin2014}
Korotin D~M, Mazurenko V~V, Anisimov V~I and Streltsov S~V 2015 {\em Phys. Rev.
  B\/} {\bf 91} 224405

\bibitem{Jacobsson2013}
Jacobsson A, Sanyal B, Le{\v{z}}ai{\'{c}} M and Bl{\"{u}}gel S 2013 {\em Phys.
  Rev. B\/} {\bf 88} 134427

\bibitem{Moreira2002}
Moreira I~P~R, Illas F and Martin R~L 2002 {\em Phys. Rev. B\/} {\bf 65} 155102

\bibitem{Archer2011}
Archer T, Pemmaraju C, Sanvito S, Franchini C, He J, Filippetti A, Delugas P,
  Puggioni D, Fiorentini V, Tiwari R and Majumdar P 2011 {\em Phys. Rev. B\/}
  {\bf 84} 115114

\bibitem{Kodderitzsch2002a}
K{\"{o}}dderitzsch D, Hergert W, Temmerman W~M, Szotek Z, Ernst A and Winter H
  2002 {\em Phys. Rev. B\/} {\bf 66} 064434

\bibitem{Faleev2004}
Faleev S~V, van Schilfgaarde M and Kotani T 2004 {\em Phys. Rev. Lett.\/} {\bf
  93} 126406

\bibitem{Kunes2007}
Kune{\v{s}} J, Anisimov V~I, Skornyakov S~L, Lukoyanov A~V and Vollhardt D 2007
  {\em Phys. Rev. Lett.\/} {\bf 99} 156404

\bibitem{Shanker1973}
Shanker R and Singh R~A 1973 {\em Phys. Rev. B\/} {\bf 7} 5000--5005

\bibitem{Hutchings1972}
Hutchings M~T and Samuelsen E~J 1972 {\em Phys. Rev. B\/} {\bf 6} 3447--3461

\bibitem{Jauch2004}
Jauch W and Reehuis M 2004 {\em Phys. Rev. B\/} {\bf 70} 195121

\bibitem{Pepy1974}
Pepy G 1974 {\em J. Phys. Chem. Solids\/} {\bf 35} 433--444

\bibitem{Rollmann2004}
Rollmann G, Rohrbach A, Entel P and Hafner J 2004 {\em Phys. Rev. B\/} {\bf 69}
  165107

\bibitem{Mazurenko2005}
Mazurenko V~V and Anisimov V~I 2005 {\em Phys. Rev. B\/} {\bf 71} 184434

\bibitem{Kren1965}
Kr{\'{e}}n E, Szab{\'{o}} P and Konczos G 1965 {\em Phys. Lett.\/} {\bf 19}
  103--104

\bibitem{Coey1971}
Coey J~M~D and Sawatzky G~A 1971 {\em J. Phys. C\/} {\bf 4} 2386--2407

\bibitem{Samuelsen1970}
Samuelsen E~J and Shirane G 1970 {\em Phys. Status Solidi\/} {\bf 42} 241--256

\bibitem{Mazurenko2007}
Mazurenko V~V, Skornyakov S~L, Kozhevnikov A~V, Mila F and Anisimov V~I 2007
  {\em Phys. Rev. B\/} {\bf 75} 224408

\bibitem{Logemann2015}
Logemann R, de~Wijs G~A, Katsnelson M~I and Kirilyuk A 2015 {\em Phys. Rev.
  B\/} {\bf 92} 144427

\end{thebibliography}

\end{document}